\documentclass{resources/aa}
\usepackage{graphicx}
\usepackage[varg]{txfonts}
\usepackage{physics}
\usepackage{hyperref}

\newcommand{\sref}[1]{Sec.~\ref{#1}}
\newcommand{\tab}[1]{Table~\ref{#1}}
\newcommand{\fig}[1]{Fig.~\ref{#1}}
\newcommand{\equ}[1]{Eq.~(\ref{#1})}
\newcommand{\equo}[1]{Eq.~\ref{#1}}
\newcommand{\equs}[2]{Eqs.~(\ref{#1})~-~(\ref{#2})}
\newcommand{\equos}[2]{Eqs.~\ref{#1}~-~\ref{#2}}
\newcommand{\Msunpyr}{\mathrm{M_\odot/yr}}

\makeatletter
\renewcommand*\aa@pageof{, page \thepage{} of \pageref*{LastPage}}
\newcommand{\pder}[2][]{\frac{\partial#1}{\partial#2}}
\makeatother

\begin{document}


%
\titlerunning{The influence of photoevaporation for protoplanetary disk evolution}
\authorrunning{M. Cecil et al.}
\title{Time-dependent long-term hydrodynamic simulations of the inner protoplanetary disk} 
\subtitle{III: The influence of photoevaporation}
%
%
\author{
 M.~Cecil\inst{1, 2} \and
 L.~Gehrig\inst{1} \and
 D.~Steiner\inst{1}
}
\institute{
 Department of Astrophysics, University of Vienna,
 Türkenschanzstrasse 17, A-1180 Vienna, Austria
 \and
Max-Planck Institute for Astronomy (MPIA), 
Königstuhl 17, 69117 Heidelberg, Germany \\
e-mail: cecil@mpia.de
}
\date{Received ....; accepted ....}

\abstract
{
The final stages of a protoplanetary disk are essential for our understanding of the formation and evolution of planets.
Photoevaporation is an important mechanism that contributes to the dispersal of an accretion disk and has significant consequences for the disk's lifetime. 
However, the combined effects of photoevaporation and star-disk interaction have not been investigated in previous studies.
}
{
A photoevaporative disk evolution model including a detailed formulation of the inner star-disk interaction region will improve the understanding of the final stages of disk evolution.  
}
{
We combined an implicit disk evolution model with a photoevaporative mass-loss profile. By including the innermost disk regions down to 0.01~AU, we could calculate the star-disk interaction, the stellar spin evolution, and the transition from an accreting disk to the propeller regime self-consistently.
Starting from an early Class~II star-disk system (with an age of 1~Myr), we calculated the long-term evolution of the system until the disk becomes almost completely dissolved.
}
{
Photoevaporation has a significant effect on disk structure and evolution. The radial extent of the dead zone decreases,
and the number of episodic accretion events (outbursts) is reduced by high stellar X-ray luminosities. 
Reasonable accretion rates ($> 10^{-8}~\mathrm{M_\odot / yr}$) in combination with photoevaporative gaps are possible for a dead zone that is still massive enough to develop episodic accretion events.
Furthermore, the stellar spin evolution during the Class~II evolution is less affected by the star-disk interaction in the case of high X-ray luminosities.
}
{
Our results suggest that the formation of planets, especially habitable planets, in the dead zone is strongly impaired in the case of strong X-ray luminosities.
Additionally, the importance of the star-disk interaction during the Class~II phase with respect to the stellar spin evolution is reduced.
}


\maketitle


\section{Introduction}
\label{sec:intro}

Planets are formed in protoplanetary disks around young stars.
To understand planet formation in these disks, we need to understand their evolution.
Over the past decades, intensive studies have been carried out to investigate the life of protoplanetary disks \citep[e.g.][]{Williams2011, Testi2014, Ercolano2017, Andrews2010, Andrews13, Andrews2020, Manara2023}.
An important part of a disk's evolution is its dispersal, which also constitutes an important benchmark for planet formation theories.
Transition disks are of particular interest, as they mark the final stage of protoplanetary evolution before being dissolved \citep[e.g.][]{Owen2016, Ercolano2017, Marel2023}.
Transition disks are characterised by a gap or cavity in the dust component close to the star starting at $\lesssim 10$~AU, with a radial extent up to $\sim 100$~AU. These disks represent approximately 10\% of the observed disk population \citep[e.g.][]{Espaillat2014, Marel2016}, suggesting that the transition phase at the end of a disk's lifetime lasts for only a few $10^5$ yr.
One particular aspect of transition disks poses a challenge in explaining their development.
Despite the gap in the disk, the majority of transition disks show accretion rates up to $\gtrsim 10^{-8}~\Msunpyr$, which are comparable to classical T~Tauri stars without observed gaps \citep[e.g.][]{Owen2016, Cieza2012, Alcala2014, Manara2017}.
The high accretion rate indicates an inner disk rich in gas and dust that coexists with the gap \citep[e.g.][]{Garate2021}.

Currently, there are two major theories explaining the mechanisms that are responsible for opening a gap in an accretion disk and retaining a high accretion rate.
First, a planet that forms in the disk can sweep up the adjacent disk material and open a gap \citep[e.g.][]{Lubow2006, Zhu2011, Zapata2017}{}{}.
Depending on the disk parameters, gas and dust can be transported across the gap and replenish the inner disk, leading to the observed high accretion rates \citep[e.g.][]{Lubow2006}.
Recent observations have indeed confirmed the presence of planets in gaps of transition disks \citep[e.g.][]{Keppler2018, Haffert2019, Currie2022}{}{}.
Unfortunately, there are observational challenges regarding the detection of additional planets in transition disks, and therefore, such detections are still pending \citep[e.g.][]{Asensio2021, Marel2021, Currie2023}.

The second promising theory for explaining the observed features of transition disks is the removal of disk material by photoevaporative winds. This approach was developed to explain the disparity between the timescales of the disk's evolution and its dispersal \citep[e.g.][]{Owen2016, Ercolano2017}. 
In this photoevaporation scenario, a wind is launched, and it removes mass from the disk in the vicinity of the gravitational radius \citep[e.g.][]{Hollenbach1994}. When combined with viscous evolution, a gap is opened in the disk structure \citep[][]{Clarke2001}, separating the disk into an inner and outer part. The inner disk is then drained onto the star quickly on a short viscous timescale, exposing the outer disk to direct irradiation from the star \citep[][]{Alexander2004}. While the two-timescale behaviour of the disk's evolution can be reproduced by this mechanism, it fails to explain the observed fraction of transition disks with large accretion rates \citep[e.g.][]{Owen2016}. To resolve this problem, the photoevaporative process can be combined with a dead zone in the inner disk, which is effectively shielded from irradiation and within which viscous transport is reduced \citep[][]{Morishima2012, Garate2021}.
While high-energy stellar irradiation opens a gap outside the dead zone, the low viscosity in the dead zone results in a long-lived inner disk that can retain accretion rates of $\sim 10^{-8}~\Msunpyr$.
Radiation in different energy regimes is capable of driving photoevaporation in disks. For instance, extreme ultraviolet (EUV) photons can remove mass at small radii, but they are easily absorbed by small column densities and typically do not cause mass-loss rates larger than $\sim 10^{-10}~\Msunpyr$ \citep[e.g.][]{Hollenbach1994, Kunitomo2020}. Far ultraviolet (FUV) radiation can penetrate larger column densities and can therefore drive a significant photoevaporative wind at radii $\gtrsim 100$~AU \citep[e.g.][]{Gorti2008, Gorti2009a}. However, the effectiveness of FUV photoevaporation (especially in the presence of other photoevaporative processes) depends highly on the disk's chemistry, such as the abundance of polycyclic aromatic hydrocarbons \citep[PAHs,][]{Wang2017, Nakatani2018}. X-ray photoevaporation, on the other hand, solely depends on the X-ray radiation properties of the star \citep[][]{Owen2010, Owen2011, Owen2012}, with a slight dependence on the stellar mass \citep[e.g.][]{Picogna2021}, and it is most effective at radii $<~100$~AU \citep[e.g.][]{Owen2010, ercolano2021}. There are several studies that have included X-ray photoevaporation and long-term viscous evolution of the disk that can recreate a wide range of gap radii and accretion rates \citep[e.g.][]{Ercolano2018, Picogna2019, Garate2021}. Other models also included the effects of a layered accretion and the occurrence of episodic accretion events \citep[e.g.][]{Bae2013PHOTO}{}{}. 
However, photoevaporative models still fail to explain certain observed features, such as the highest observed accretion rates in transition disks \citep[$\sim 10^{-7}~\Msunpyr$, e.g.,][]{Marel2023}.
Given the characteristics of both gap-opening mechanisms, it is most likely that the whole distribution of transition disks can be explained by a combination of the aforementioned (and possibly other) processes.

In this study, we focus on the hydrodynamic disk evolution with the implicit TAPIR code \citep[e.g.][]{Ragossnig2019, Steiner21, Gehrig2022, Gehrig2023Paper2} in combination with the mass-loss profile due to X-ray photoevaporation presented in \cite{ercolano2021}.
The influence of a planetary companion is not considered in this work and will be added to our model in the future.
In our model, we can include the innermost disk region up to 0.01~AU \citep[][]{Steiner21}{}{}.
In this important region close to the star, the star and disk exchange mass and angular momentum via the connection by magnetic field lines \citep[e.g.][]{Koenigl91, Shu94, Romanova09, hartmann16}, which also affects the stellar spin evolution \citep[e.g.][]{Matt10, Gallet19, Ireland21}.
Additionally, the balance between the stellar magnetic pressure and the disk pressure defines the position of the inner disk boundary \citep[e.g.][]{hartmann16}.
Our model also incorporates the layered disk model (active layer and dead zone) introduced by \cite{gammie96} and the possibility of episodic accretion events (outbursts) due to magneto-rotational instabilities \citep[MRI,][]{Balbus1991}.

The aim of this paper is to combine a detailed model of the inner disk and a photoevaporative profile in long-term calculations.
We start at early Class~II star-disk systems with an age of 1~Myr and simulate their evolution over several million years until the accretion rate drops towards zero.
With our model, we can calculate the combined effects of the inner disk boundary, the magnetic star-disk interaction, the stellar spin evolution, and the photoevaporative mass-loss, and we can work out the back-reaction between the disk and the star.
The paper is structured as follows: \sref{sec:model_description} contains the disk, photoevaporative, and stellar spin model.
Our results are presented in \sref{sec:results} and put into context with respect to previous works, observations, and implications for the formation and evolution of planets in \sref{sec:Discussion}.
Finally, our conclusions are summarised in \sref{sec:conclusion}.

\section{Model description}
\label{sec:model_description}

We describe the structure and the evolution of protoplanetary disks by combining the equations of hydrodynamics with a turbulent viscosity description \citep[e.g.][]{shakura73}.
The disk model is summarised in \sref{sec:hydrodynamic_disk_evolution}, \sref{sec:boundaries}, and \cite{Steiner21}.
The photoevaporative model and the stellar spin model are highlighted in \sref{sec:PE} and \sref{sec:stellar_spin_model}, respectively.
Finally, we summarize our model parameters in \sref{sec:model_parameters}.

\subsection{Hydrodynamic disk evolution}\label{sec:hydrodynamic_disk_evolution}

In this study, long-term global disk simulations were carried out with the implicit hydrodynamic TAPIR code \citep[e.g.][]{Ragossnig2019, Steiner21}.
Without resorting to the diffusion approach \citep[e.g.][]{pringle81, Armitage01}, our model can include deviations from a Keplerian angular velocity, the influence of stellar magnetic torques, and pressure gradients \citep[e.g.][]{Steiner21}.
These effects are significant in the innermost region of the disk \citep[e.g.][]{Romanova02, bessolaz08}.
In the current version of the TAPIR code, an axisymmetric disk is assumed, which is in hydrostatic equilibrium in the vertical direction.
The protoplanetary disk in our simulations is described in \equs{eq:cont}{eq:ene} as a time-dependent, vertically integrated, viscous accretion disk \citep[e.g.][]{shakura73, Armitage01},
\begin{alignat}{2}
    & \pder{t} \, \Sigma &&+ \nabla \cdot ( \Sigma \, \vec u ) + \Dot{\Sigma}_\mathrm{PE} = 0\;, \label{eq:cont} \\
    & \pder{t} (\Sigma \, \vec u) &&+ \nabla \cdot (\Sigma \, \vec u : \vec u) - \frac{B_\mathrm{z} \vec B}{2 \pi}\nonumber \\
    & &&+ \nabla P_\mathrm{gas} + \nabla \cdot Q + \Sigma \, \nabla \psi + H_\mathrm{P} \nabla \left( \frac{B_\mathrm{z}^2}{4 \pi} \right) + \Dot{\Sigma}_\mathrm{PE} \vec u = 0 \;, \label{eq:mot} \\
    &\pder{t} (\Sigma \, e) &&+ \nabla \cdot (\Sigma \, \vec u \, e ) + P_\mathrm{gas} \, \nabla \cdot  \vec u \nonumber \\
    & &&+ Q : \nabla \vec u - 4 \pi \, \Sigma \, \kappa_\mathrm{R}\left(J - S \right) + \dot E_\mathrm{rad} + \Dot{\Sigma}_\mathrm{PE} e = 0 \;, \label{eq:ene}
\end{alignat}
where $\Sigma$, $\vec u$, $e$, $P_\mathrm{gas}$ and $H_\mathrm{P}$ denote the gas column density, gas velocity in the planar components $\vec u = (u_\mathrm{r}, u_\mathrm{\varphi})$, specific internal energy density, vertically integrated gas pressure and the vertical scale height of the gas disk, respectively.
The mass loss due to photoevaporation, $\Dot{\Sigma}_\mathrm{PE}$, is presented in further detail in \sref{sec:PE}.
The gradient in planar cylindrical coordinates reads $\nabla = (\partial / \partial r , r^{-1} \partial / \partial \varphi)$ with $\partial / \partial \varphi = 0$ for the axisymmetric model.
$P_\mathrm{gas}$ is defined by the ideal equation of state $P_\mathrm{gas} = \Sigma e ( \gamma - 1 )$, with the adiabatic coefficient $\gamma = 5 / 3$ \citep[e.g.][]{bessolaz08} and $\kappa_\mathrm{R}$ denotes the Rosseland-mean opacity \citep[e.g.][]{Mihalas84}, which is composed of a gaseous component $\kappa_\mathrm{R,gas}$ \citep[based on][]{caffau11} and a dust-dominated component $\kappa_\mathrm{R,dust}$ \citep[based on][]{pollack85}; with $\kappa_\mathrm{R} = \kappa_\mathrm{R,gas} + f_\mathrm{dust} \kappa_\mathrm{R,dust}$, where $f_\mathrm{dust} = 0.014$ is the dust-to-gas mass ratio based on solar metallicity.
The gravitational potential of the star-disk system is denoted by $\psi$.
Viscous angular momentum transport and viscous heating are represented by the viscous stress tensor $Q$ in their corresponding terms in \equ{eq:mot} and \equ{eq:ene}, respectively.
The viscosity, $\nu$, in our model follows \cite{shakura73}, 
\begin{equation}
    \nu = \alpha \, c_\mathrm{S} \,  H_\mathrm{P} \;, \label{eq:shakura} \\ 
\end{equation}
with the viscosity parameter $\alpha$ and the speed of sound, $c_\mathrm{S}$.
To include the effects of a dead zone, we used the layered disk model \citep[][]{gammie96}{}{}.
The disk can be divided into two layers.
A sufficiently ionised, MRI active layer where angular momentum and mass can be transported effectively and a layer that is sufficiently shielded from stellar and external radiation (the dead zone), in which radial transport of mass and angular momentum is reduced.
In this context, we assumed that an upper (surface) layer of the disk with $\Sigma_\mathrm{surf} = 100~\mathrm{g/cm^2}$ is always sufficiently ionised and MRI active \citep[e.g.][]{Vorobyov20}{}{}.
The viscous parameters in the MRI active layer ($\alpha_\mathrm{MRI}$) and the dead zone ($\alpha_\mathrm{DZ}$) were taken from \cite{Steiner21} and \cite{Gehrig2022}.
When the temperature in the midplane reaches an activation temperature of $T_\mathrm{active} \sim 1000$~K, the viscous $\alpha$ value increases from $\alpha_\mathrm{DZ}$ to $\alpha_\mathrm{MRI}$ \citep[see implementation in][]{Steiner21}{}{}.
The heating and cooling rate per unit surface area, $\Dot{E}_\mathrm{rad}$, includes stellar irradiation, surface cooling, and heating from an ambient medium with a temperature of 20~K.
The radiative transport ($(J - S)$-term in \equ{eq:ene}) was approximated by a diffusion approximation with an Eddington factor of 1/3 \citep[e.g.][]{Ragossnig2019, Steiner21}.

In addition, we took into account the influence of a magnetic field $\vec B = (B_\mathrm{z}, B_\mathrm{\varphi})$ on the disk, which is approximated as a dipole.
The magnetic field was assumed to rotate rigidly with the stellar angular velocity, $\Omega_\star$.
The vertical component of the magnetic field $B_\mathrm{z}$ remains constant within the vertical range of the disk, but it can vary radially. 
As the stellar magnetic field lines are dragged in angular direction by the disk material, the field lines wind up and the angular component of the magnetic field, $B_\mathrm{\varphi}$, depends on the ratio of the stellar angular velocity and the local angular disk velocity, $\Omega_\star / \Omega(r)$ \citep[e.g.][]{rappaport04, kluzniak07, Steiner21}.

Finally, our model solved an adaptive grid equation together with the hydrodynamic equations to provide sufficient radial grid point resolution \citep[e.g.][]{Dorfi1987, Ragossnig2019}. 
The adaptive grid equation also allows the model to adapt the location of the inner and outer disk boundaries according to the disk evolution (see \sref{sec:boundaries}).

\subsection{Disk boundaries}\label{sec:boundaries}

The inner disk boundary of our model was motivated by the equilibrium of the magnetic pressure from the stellar magnetic field, $P_\mathrm{mag}$, and the pressure of the disk material that contains the ram pressure of the in-falling material, $P_\mathrm{ram}$, and the gas pressure, $P_\mathrm{gas}$.
At the truncation radius, $r_\mathrm{t}$, $P_\mathrm{mag}$ equals the pressure contribution that dominates in the disk \citep[e.g.][]{Koldoba02,Romanova02,bessolaz08},
\begin{equation}\label{eq:r_trunc}
    P_\mathrm{magn}(r_\mathrm{t}) = \text{max}\left[P_\mathrm{ram}(r_\mathrm{t}), P_\mathrm{gas}(r_\mathrm{t})\right]\;.
\end{equation}
In case the ram pressure dominates the local gas pressure in the disk, the truncation radius yields \citep[][]{hartmann16}{}{},
\begin{align}
    r_{\mathrm{t}}(P_\mathrm{ram} \geq P_\mathrm{gas}) \approx 18 \, \xi \, R_\odot & \, \left(\frac{B_\star}{10^3 \, G}\right)^{4/7}  \left(\frac{R_\star}{2 \, R_\odot}\right)^{12/7} \left(\frac{M_\star}{0.5 \, M_\odot}\right)^{-1/7} \nonumber \\ 
    & \left(\frac{\dot M_\star}{10^{-8} \, M_\odot / \mathrm{yr}}\right)^{-2/7} \label{eq:truncation_radius} \;, 
\end{align}
with the stellar magnetic field strength, $B_\star$, the stellar radius, $R_\star$, the stellar mass, $M_\star$, the accretion rate on the star, $\Dot{M}_\star$, and a correction factor $\xi = 0.7$ \citep[][]{hartmann16}{}{}.
For $P_\mathrm{ram} < P_\mathrm{gas}$, the truncation radius follows from equating the magnetic pressure of the vertical field component $B_\mathrm{z}$ with the gas pressure.
The outer disk boundary was set to the radius where $\Sigma = 1.0$~$\mathrm{g/cm^2}$ \citep[similar to][]{Vorobyov20}.

For the hydrodynamic equations \equs{eq:cont}{eq:ene} the numerical boundary conditions were chosen as "free" (Neumann) boundary conditions 
(e.g. $\partial \Sigma / \partial r = 0$) at the inner boundary \citep[e.g.][]{Romanova02,Romanova04}. 
At the outer disk boundary, the angular and radial velocities were fixed to the Keplerian velocity and zero, respectively.
The surface density and the specific internal energy were treated with "free" boundary conditions at the outer disk boundary.

\subsection{Photoevaporation model}\label{sec:PE}

The influence of photoevaporation on the evolution of our disk model was represented as sink terms in the system of hydrodynamic equations (\equs{eq:cont}{eq:ene}). The values of $\dot \Sigma_{\mathrm{PE}}$ at every point throughout the extent of the disk were derived from the description of \cite{ercolano2021}, which is an extension of the model created by \cite{Picogna2019}. Here we briefly lay out the structure of the photoevaporative mass-loss term, following \cite{ercolano2021}. The 1D radial profile of the photoevaporative mass-loss rate is only dependent on the radial distance from the star $r$ and the stellar X-ray luminosity in the "soft" part of the X-ray spectrum ($h\nu$< 1 keV) $L_\mathrm{X,soft}$ and is described by,

\begin{multline}
    \label{eq:photoprofile}
    \dot \Sigma_{\mathrm{PE}} (r)= \chi \mathrm{ln}(10) \Biggl( \frac{6a\mathrm{ln}(r)^5}{\mathrm{ln}(10)^6} + \frac{5b\mathrm{ln}(r)^4}{\mathrm{ln}(10)^5} + \frac{4c\mathrm{ln}(r)^3}{\mathrm{ln}(10)^4} + 
    \\ \frac{3d\mathrm{ln}(r)^2}{\mathrm{ln}(10)^3} + 
    \frac{2e\mathrm{ln}(r)}{\mathrm{ln}(10)^2} +
    \frac{f}{\mathrm{ln}(10)}\Biggr) \frac{\dot M_{\mathrm{PE}}(r)}{2 \pi r^2} 
    \mathrm{exp} \left[ -\left( \frac{r}{r_\mathrm{cut}}\right)^{10} \right]\;,
\end{multline}

where $\chi$ is a unit conversion factor with the value of 0.2817 to convert from $[\mathrm{M}_\odot~\mathrm{au}^{-2}~\mathrm{yr}^{-1}]$ to $[\mathrm{g~cm}^{-2}~\mathrm{s}^{-1}]$, $r_\mathrm{cut}$ is the cutoff radius beyond which the mass-loss is suppressed and $\dot M_{\mathrm{PE}}(r)$ is the cumulative mass-loss rate,

\begin{multline}
    \label{eq:cumulmass}
    \mathrm{log}_{10}\Biggl( \frac{\dot M_{\mathrm{PE}}(r)}{\dot M (L_{\mathrm{X, soft}})} \Biggr)=a \mathrm{log}(r)^6+ b\mathrm{log}(r)^5+c\mathrm{log}(r)^4+\\
    d\mathrm{log}(r)^3+e\mathrm{log}(r)^2+f\mathrm{log}(r)+g \;.
\end{multline}

The dependence of the mass-loss rate with respect to the (soft part of the) stellar X-ray luminosity is given by, 

\begin{equation} \label{eq:photoXraydependence}
     \mathrm{log}(\dot M(L_{\mathrm{X,soft}})) = A_{\mathrm{L}} \mathrm{exp} \left [ \frac{(\mathrm{ln}(\mathrm{log}(L_{\mathrm{X,soft}}))-B_{\mathrm{L}})^2}{C_{\mathrm{L}}}\right] + D_{\mathrm{L}} \;.
\end{equation} 

\begin{figure}[ht!]
    \centering
         \resizebox{\hsize}{!}{\includegraphics{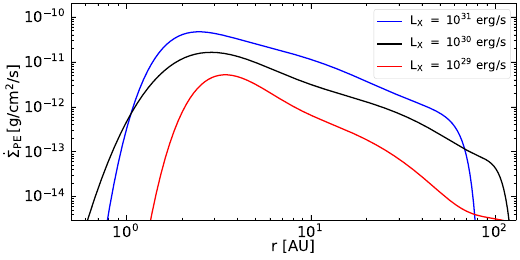}}
    \caption{
    Mass-loss rate due to photoevaporation for different X-ray luminosities with respect to the disk radius. 
    The blue, black, and red lines correspond to an X-ray luminosity of $L_\mathrm{X} = 10^{31}$~erg/s, $10^{30}$~erg/s, and $10^{29}$~erg/s, respectively.
    }
    \label{fig:sigmadot}
\end{figure}

\cite{ercolano2021} used three different synthesised input spectra for their calculations of the 1D mass-loss profiles. They correspond to stellar X-ray luminosities of $L_\mathrm{X} = 10^{31}$~erg/s, $10^{30}$~erg/s, and $10^{29}$~erg/s and increase in "hardness" for larger luminosities. The "hardness" of a spectrum is hereby defined as the fraction of the integrated soft part of the X-ray spectrum ($0.1~\mathrm{keV} \leq h\nu \leq 1$  keV) to the total integrated X-ray flux. As already found by \cite{Ercolano2009}, the soft X-rays dominate the heating mechanism and are therefore the main driver of X-ray photoevaporation, while the flux of hard X-rays only plays a minor role. This is supported by findings that show no significant X-ray-induced photoevaporative mass-loss when the used stellar X-ray spectrum is dominated by the high energy regime \citep[e.g.][]{Wang2017, Nakatani2018}. For that reason, \cite{ercolano2021} argue that it is sufficient to only consider the soft part of the X-ray spectra ($L_\mathrm{X,soft}$) to determine accurate mass-loss rates. Following Tab. 4 of \cite{ercolano2021}, the soft X-ray flux makes up 52\% of the total X-ray luminosity in the spectrum corresponding to $10^{29}$ erg/s and this percentage decreases to 47\% and further to 42\% for the $10^{30}$ erg/s and $10^{31}$ erg/s spectra, respectively. The coefficients $a~-~f$ as well as the cutoff radius $r_\mathrm{cut}$ depend on the considered X-ray spectrum while $A_\mathrm{L}~-~D_\mathrm{L}$ are constant for all used input spectra. The numerical values for $a~-~f$, $r_\mathrm{cut}$ and $A_\mathrm{L}~-~D_\mathrm{L}$ are given in Tab. 3 and section 3.3, respectively, of \cite{ercolano2021}. The resulting radial photoevaporative mass-loss profiles for the three different spectra and corresponding X-ray luminosities are depicted in \fig{fig:sigmadot}.\\
As shown in \equs{eq:cont}{eq:ene}, the sink term describing the mass loss due to photoevaporation does not only enter the equation of continuity (\equ{eq:cont}) but also has to be considered in the equations of motion (\equ{eq:mot} and internal energy (\equ{eq:ene}). In these last two equations, the terms $\dot \Sigma_\mathrm{PE} \vec u$ and $\dot \Sigma_\mathrm{PE} e$ represent the momentum and internal energy, respectively, which is carried by the photoevaporated particles and consequently extracted from the disk as well. If these terms were not considered, the extracted mass would leave behind a surplus of momentum and internal energy, which would effectively lead to an unphysical acceleration and heating of the surrounding disk material.
We illustrate these effects in Appendix \ref{Appendix}.


\subsection{Stellar spin model}\label{sec:stellar_spin_model}

To model the stellar spin evolution, we used the same model as described in \cite{Gehrig2022}.
The star rotates as a rigid, fully convective body \citep[e.g.][]{Armitage96, Matt10, Matt12, Pantolmos20} that contracts along the Hayashi track \citep[e.g.][]{Siess00, Collier93, Matt10}.
The stellar angular momentum changes according to external torques acting on the star. 
Over the past decade, several studies have quantified the external torques based on stellar and disk parameters \citep[e.e.g,][]{Matt10, Gallet19, Ireland21}.

Following the model presented in \cite{Gallet19}, the accretion of material adds angular momentum to the star and results in a stellar spin-up. 
The accretion process can launch an accretion-powered stellar wind \citep[APSW][]{Matt05b} that ejects material away from the star and consequently causes a spin-down.
The amount of angular momentum removed by the APSW depends on the amount of ejected material, which scales with the accretion rate on the star multiplied by a factor, $W \lesssim 1~\%$ \citep[e.g.][]{Crammer08, Pantolmos20}{}{}.
Finally, the star-disk interaction can transfer angular momentum between the disk and the star in the form of magnetospheric ejections \citep[MEs; e.g.][]{Zanni13, Gallet13}.
The amount and sign of the transferred angular momentum depend on the differential rotation between the stellar magnetic field lines and the disk material as well as on the position of the truncation radius with respect to the co-rotation radius,
\begin{equation}
    r_\mathrm{cor} = \left( \frac{G M_\star}{\Omega_\star^2} \right)^{1/3} \, ,
\end{equation}
with the gravitational constant, $G$.
For a more detailed description of the spin model, we kindly refer to \cite{Gallet19} and \cite{Gehrig2022}.

\subsection{Initial model} \label{sec:init}
The construction of the stationary initial model followed the description given by \cite{Steiner21}. 
Here we summarise the main points of its derivation. \\
The initial model for our simulations was a stationary solution to the set of equations given by Eqs. \ref{eq:cont}~-~\ref{eq:ene} with the last terms on the left-hand side (corresponding to the photoevaporative process) set to 0. We started the construction of this model by first adopting a simple Keplerian disk with no velocity component in the radial direction and then introducing a mass flux into the disk across the outer boundary and conducting the time integration of the hydrodynamic equations together with the adaptive grid equation. The disk structure adjusts to this mass flux according to the adapted boundary conditions, stellar parameters, opacities, stellar magnetic torques in the inner disk and viscosity description. In the stationary state, the mass flux throughout the whole disk (and consequently the accretion rate on the star) is the same as the one introduced across the outer boundary. We refer to this value of the mass flux as $\dot{M}_\mathrm{init}$. 

\subsection{Model parameters}\label{sec:model_parameters}

We constructed the model in this study similar to \cite{Gehrig2022} and followed the evolution of a disk around a $0.7~\mathrm{M_\odot}$ T~Tauri (Class~II) star.
The initial age of the star was $t_\mathrm{0}=1$~Myr and the initial stellar radius and the effective temperature were taken from \cite{Baraffe15}.
The other stellar and disk parameters are summarised in \tab{tab:model_para}.
In case we deviate from these values, we mark the respective changes.
With these parameters, we constructed an initial steady-state model following \cite{Steiner21}.
The initial disk model has a mass of $2~\%$ of the stellar mass and an outer radius of 127~AU.
The disk mass and radial extension coincide with observations of young star-forming regions \citep[e.g.][]{Andrews2010, Ansdell2018}{}{}.

\begin{table}[ht]
\centering
\caption{Model parameters used in this study.}              
\begin{tabular}{l r | l r}         
\hline\hline 
$B_\star$ [kG] & 1.00 & $R_\mathrm{\star,~init}$ [$R_\odot$] & 2.096 \\

$P_\mathrm{init}$ [days] & 6.0 & $M_{\star,~\mathrm{init}}$ [$\mathrm{M_\odot}$] & 0.7 \\

$\Dot{M}_\mathrm{init}$ [$\mathrm{M_\odot/yr}$] & $10^{-8}$ & $T_\mathrm{eff}$ [K] & 4078  \\

$W$ [\%] & 1 & $T_\mathrm{active}$~[K] & 1000 \\

$\alpha_\mathrm{MRI}$ & 0.02 & $\alpha_\mathrm{DZ}$ & 0.0004 \\

\hline\hline                                            
\end{tabular}
\label{tab:model_para}  
\end{table}


\section{Results}
\label{sec:results}

\subsection{Impact of photoevaporation on the disk structure}\label{sec:disk_evo}

\begin{figure}[ht!]
    \centering
         \resizebox{\hsize}{!}{\includegraphics{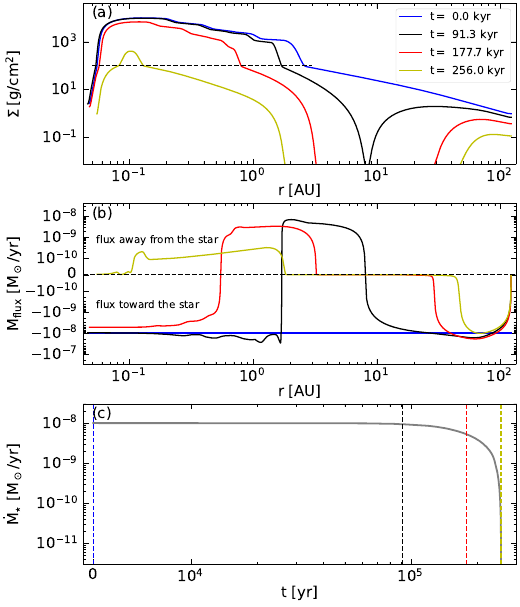}}
    \caption{
    Exemplary evolution of our model. 
    The X-ray luminosity of the star is $L_\mathrm{X}=10^{31}~\mathrm{erg/s}$.
    Panel~a shows the radial surface density profile, $\Sigma(r)$, of four different timesteps. 
    The initial model is shown as a blue line.
    From the opening of the gap (black line), we follow the evolution until the accretion rate on the star drops to zero (yellow line).
    The horizontal dashed line shows the surface layer density of $\Sigma_\mathrm{surf}=100~\mathrm{g/cm^2}$.
    We assumed the presence of a dead zone if the surface density of the disk is larger than $\Sigma_\mathrm{surf}$.
    Panel~b shows the mass flux within the disk for each timestep.
    A negative value denotes a flux towards the star, and a positive value corresponds to a mass flux away from the star.
    The horizontal dashed line shows the value of zero and divides the flux towards and away from the star.
    Panel~c shows the evolution of the accretion rate, $\Dot{M}_\star$, over time.
    The coloured dashed vertical lines correspond to the respective surface density profile in Panel~a.
    }
    \label{fig:ref_mod}
\end{figure}

We want to start with the evolution of an illustrative model showing the effect of photoevaporation.
The X-ray luminosity of the star for this simulation was $L_\mathrm{X} = 10^{31}~\mathrm{erg/s}$, which corresponds to a soft X-ray luminosity $L_\mathrm{X,soft}=4.2\cdot10^{30}~\mathrm{erg/s}$ (see \sref{sec:PE}).
The surface density profile of the initial model is shown in Panel~a of \fig{fig:ref_mod}.
The shape of this profile was determined as described in Sec. \ref{sec:init} and \cite{Steiner21}. 
Panel~b shows the mass flux within the disk, which is radially constant with a value of $10^{-8}~\Msunpyr$ for our initial model.
The evolution of the accretion rate on the star is shown in Panel~c.
We note that we `switched off' the occurrence of episodic accretion events for this example.
In this context, the activation temperature, $T_\mathrm{active}$, is increased so that accretion events cannot develop.
This serves the purpose of illustration and saves computational resources.

The formation of a gap in the surface density structure of the disk is a consequence of the quenching of inward radial mass flux due to photoevaporation: As the disk material flows towards the star from the outer disk, the mass in the flow is continually diminished by the removal of matter by photoevaporation. 
In the dead zone, the (initially radially constant) mass flux is manifested by larger amounts of mass moving at lower radial velocities compared to the area in the outer disk beyond the dead zone \citep[e.g.][]{Steiner21}. This means photoevaporation only has a very small effect on the flux in the dead zone relative to the outer disk.
In the early stages of evolution, the mass flux in the outer disk is large enough to compensate for the mass loss due to photoevaporative winds before it reaches the dead zone.
Viscous evolution causes the flux in the outer disk to be, on average, decreased as time progresses, which means the depletion of the disk mass due to photoevaporation becomes increasingly noticeable. Eventually, the flux has decreased to an extent that photoevaporation is able to reduce it to zero before the material reaches the dead zone. The location in the disk where this is the case at the earliest time is the outer edge of the dead zone since the mass flux can be kept at high values at smaller radii due to the large density and consequent relatively weak impact of photoevaporative mass-loss on the flux. As the inward mass flux in the outer disk continues to decrease, the point where photoevaporation diminishes the flux to zero moves outwards. In the region between this point and the outer edge of the dead zone, the flux becomes inverted, meaning that the disk material is moving away from the star due to a negative pressure gradient and outward angular momentum transport from the inner disk. The outer boundary of the flux inversion zone is a location where the impact of photoevaporative mass removal on the surface density structure is very effective because there is no resupply of material coming from either the inner or the outer disk. As soon as this point reaches a radius where the surface density is sufficiently small for the local photoevaporative mass loss to reduce the surface density to zero before the flux inversion zone can be pushed further, a gap is created.

In the model shown in \fig{fig:ref_mod}, this happens at a time of $t=91.3$~kyr (black lines). 
Up to this point in time, the accretion rate onto the star has only slightly decreased compared to its initial value of $\Dot{M}_\mathrm{init} = 10^{-8}~\Msunpyr$. As the gap widens, the accretion rate diminishes quickly and the dead zone becomes smaller (red lines). In Panel~a of \fig{fig:ref_mod}, the threshold of the dead zone is marked as a horizontal dashed line. Meanwhile, the inner border of the flux inversion region stays at the outer rim of the dead zone and moves towards the inner disk boundary alongside the shrinking dead zone. As the accretion rate decreases, the magnetic pressure pushes the inner disk radius further out towards the co-rotation radius (see \equo{eq:truncation_radius}).
As shown by the yellow line in Panel~b, corresponding to a time of $t=256.0$~kyr, the inner rim of the flux inversion region is able to cross the dead zone and arrive at the inner boundary of the disk. This is possible due to the stellar magnetic torque acting on the inner disk: Outside the co-rotation radius, the stellar magnetosphere moves at a greater angular velocity than the disk material, therefore exerting a positive torque on the disk and inducing angular momentum that is transported outwards by viscous interaction. This surplus of angular momentum reaches the inner edge of the flux inversion zone, accelerating the material in an azimuthal direction at that location and pushing it away from the star. This effectively leads to an inwards movement of the flux inversion zone. As the dead zone becomes smaller, the effective viscosity decreases, and the radially inward flux diminishes until the angular momentum induced by the stellar magnetic torque cannot be countered anymore, and the flux inversion region can expand towards the inner disk boundary. At this point in time which we denote with $\tau_\mathrm{end}$ (corresponding to $t=256.0$~kyr for the simulation in \fig{fig:ref_mod}), the inner boundary is pushed over the co-rotation radius, the dead zone has an extension of $\lesssim 0.1$~AU and the accretion of material on the star comes to a halt (yellow lines in \fig{fig:ref_mod}). We stopped our simulations at $\tau_\mathrm{end}$.

The remaining inner disk enters the propelling regime \citep[e.g.][]{Ustyugova2006} and is most likely dissolved by a combination of magneto-centrifugal winds \citep[e.g.][]{Konigl11, Bai2013, Kunitomo2020, Lesur2022} and photoevaporation. 
The influence of a disk in the propelling regime on the star and vice versa will be studied in detail in a subsequent work (Cecil~et~al. in~prep.)
The outer disk will disperse due to internal or external photoevaporation but should not affect the inner disk or the stellar evolution significantly after the gap has developed.

\begin{figure}[ht!]
    \centering
         \resizebox{\hsize}{!}{\includegraphics{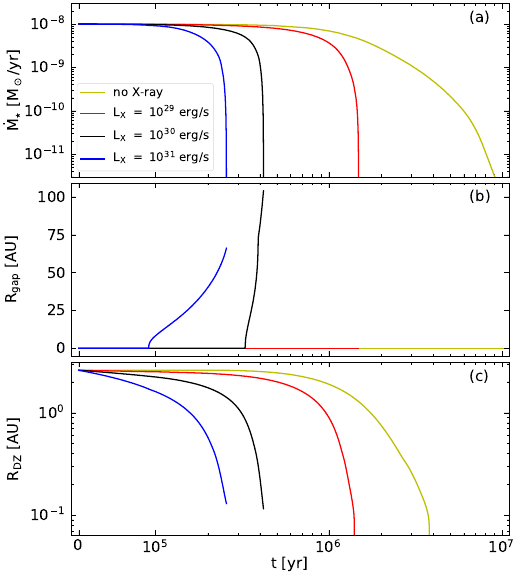}}
    \caption{
    Disk evolution for different X-ray luminosities. The blue, black, red, and yellow lines correspond to an X-ray luminosity of $L_\mathrm{X} = 10^{31}~\mathrm{erg/s}$, $10^{30}~\mathrm{erg/s}$, $10^{29}~\mathrm{erg/s}$, and no X-ray luminosity, respectively.
    Panel~a,~b, and c show the mass accretion rate on the star, $\Dot{M}_\star$; the width of the photoevaporative gap, $R_\mathrm{gap}$; and the outer edge of the dead zone, $R_\mathrm{DZ}$, respectively.
    }
    \label{fig:time}
\end{figure}

The effect of photoevaporation originating from different X-ray luminosities is illustrated in \fig{fig:time}.
The blue, black, red, and yellow lines correspond to an X-ray luminosity of $L_\mathrm{X} = 10^{31}~\mathrm{erg/s}$, $10^{30}~\mathrm{erg/s}$, $10^{29}~\mathrm{erg/s}$, and no X-ray luminosity, respectively.
Starting from an identical initial model, the disk lifetime is significantly affected by the different intensities of photoevaporation.
Without the effects of X-ray photoevaporative mass-loss, the disk lasts over 10~Myr ($\tau_\mathrm{end} = 10.1$~Myr).
Assuming a strong X-ray luminosity (blue lines), the disk dissolves over an order of magnitude faster ($\tau_\mathrm{end}= 0.256$~Myr).
Even if no gap develops (red lines), a weak X-ray luminosity can reduce the disk lifetime by a factor of $\sim 6$, compared to the case with no photoevaporation.
The outer edge of the dead zone, $R_\mathrm{DZ}$, is also affected by the strength of the X-ray luminosity.
In case of a strong X-ray luminosity of $L_\mathrm{X}= 10^{31}~\mathrm{erg/s}$ (blue line), $R_\mathrm{DZ}$ decreases below 1~AU at a time of 0.16~Myr.
Without photoevaporation (yellow line), this point in time is reached at 1.73~Myr.
Finally, we define the time in which the gap has been opened as $\tau_\mathrm{gap}$ and calculate the ratio between the lifetime of the disk after the gap has been opened and the total disk lifetime, $f_\mathrm{gap} = (\tau_\mathrm{end}~-~\tau_\mathrm{gap}) / (\tau_\mathrm{end}~+~t_\mathrm{0})$.
For X-ray luminosities of $10^{31}~\mathrm{erg/s}$ and $10^{30}~\mathrm{erg/s}$, we find $f_\mathrm{gap} = 0.13$ and $f_\mathrm{gap} = 0.06$, respectively.
We note that these values are likely lower boundaries as we stopped our simulations at the transition to the propeller regime.
When including the final dispersal of the disk, the values of $f_\mathrm{gap}$ could increase slightly. 
Compared to the observed fraction of transition disks of $\sim 10\%$ \citep[e.g.][]{Kenyon1995, Ercolano2017, Marel2023}, the values of $f_\mathrm{gap}$ are in reasonable agreement when considering that not all stars have X-ray luminosities $> 10^{30}~\mathrm{erg/s}$ and that there are other gap opening mechanisms like planetary companions.

\subsection{Photoevaporation and episodic accretion events}\label{sec:burst_evo}

To study the effects of photoevaporation on the disk evolution including episodic accretion events, we set the MRI activation temperature, $T_\mathrm{active}$, to 1000~K.
If the midplane temperature in the dead zone reaches $T_\mathrm{active}=1000$~K, the viscosity parameter increases from $\alpha_\mathrm{DZ}= 0.0004$ to $\alpha_\mathrm{MRI}= 0.02$ (see \tab{tab:model_para}) and a phase of enhanced accretion is triggered \citep[for a detailed description of these events see, e.g.,][and references therein]{Steiner21}.

\begin{figure}[ht!]
    \centering
         \resizebox{\hsize}{!}{\includegraphics{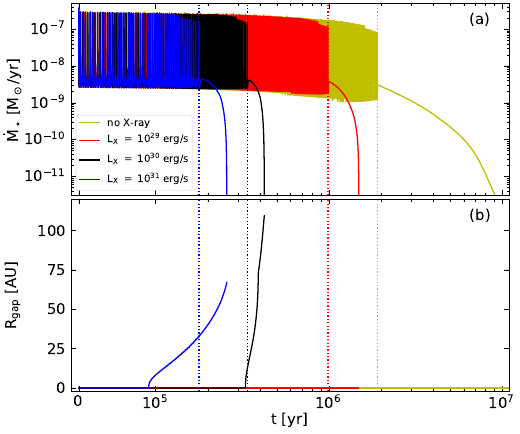}}
    \caption{
    Disk evolution for different X-ray luminosities including episodic accretion events. The different coloured lines correspond to the same X-ray luminosities in \fig{fig:time}.
    Panel~a and b show the mass accretion rate on the star, $\Dot{M}_\star$, and the width of the photoevaporative gap, $R_\mathrm{gap}$, respectively.
    The vertical coloured lines show the end of the episodic accretion events, $\tau_\mathrm{eae}$.
    }
    \label{fig:timeBurst}
\end{figure}

In \fig{fig:timeBurst}, the evolution of the accretion rate on the star, $\Dot{M}_\star$, and the width of the photoevaporative gap, $R_\mathrm{gap}$, is shown.
With different X-ray luminosities, the number and total duration of the accretion events vary significantly.
In this context, we define the point in time after which no more episodic accretion events are occurring as $\tau_\mathrm{eae}$.
The characteristics of the episodic accretion events for the different X-ray luminosities are summarised in \tab{tab:burst_chara}.
Similar to the disk lifetime, a strong X-ray luminosity reduces the time span, in which episodic accretion events are triggered.
Consequently, the number of these events also decreases with increasing X-ray luminosity.
Without photoevaporation, a total of 660 accretion events occur over a time span of 1.90~Myr.
For the model with $L_\mathrm{X} = 10^{31}~\mathrm{erg/s}$, only 55 accretion events occur over a time span of 0.18~Myr.

Interestingly, the two models with a high X-ray luminosity show episodic accretion events after the gap has formed (see the vertical black and blue lines in \fig{fig:timeBurst}).
Although the flow of material from the outer disk regions towards the dead zone is interrupted due to the photoevaporative gap, enough mass is stored in the dead zone to trigger several episodic accretion events. 
The gap radius at $\tau_\mathrm{eae}$ and the relative time fraction, in which an accretion event is active after the gap has formed, $f_\mathrm{eae}$, is also summarised in \tab{tab:burst_chara}.
In this context, we defined an accretion event as active when the accretion rate on the star exceeds $10^{-8}~\Msunpyr$ and the accumulated time in which the accretion event is active after the gap has formed was defined as $\tau_\mathrm{accum}$.
Thus, we could calculate $f_\mathrm{eae} = \tau_\mathrm{accum}/ (\tau_\mathrm{end}-\tau_\mathrm{gap})$.
The gap radii at $\tau_\mathrm{eae}$ range between 11.8~AU and 32.7~AU for $L_\mathrm{X} = 10^{30}~\mathrm{erg/s}$ and $L_\mathrm{X} = 10^{31}~\mathrm{erg/s}$, respectively. 
The value of $f_\mathrm{eae}$ is of the order of a few percent for both X-ray luminosities.

The sensitivity of $R_\mathrm{gap}(\tau_\mathrm{eae})$ and $f_\mathrm{eae}$ were then tested for different model parameters.
In the model `$\mathrm{low\_alpha}$', we reduced the value of $\alpha_\mathrm{MRI}$ to 0.005 and the initial accretion rate to $\Dot{M}_\mathrm{init} = 5.0\times 10^{-9}~\Msunpyr$.
In the model `$\mathrm{low\_Bstar}$', we reduced the stellar magnetic field strength to 0.5~kG.
All other parameters remained unchanged and the X-ray luminosity was set to $10^{31}~\mathrm{erg/s}$.

\begin{figure}[ht!]
    \centering
         \resizebox{\hsize}{!}{\includegraphics{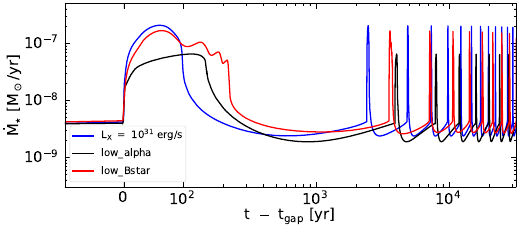}}
    \caption{
    Accretion rate on the star over time for the models `$\mathrm{low\_alpha}$' (black line) and `$\mathrm{low\_Bstar}$' (red line) and the reference model (blue line). 
    The X-ray luminosity for all models is set to $10^{31}~\mathrm{erg/s}$.
    The time axis is shifted to the beginning of the first accretion event after the gap has formed. 
    }
    \label{fig:burstcomp}
\end{figure}

Both models trigger episodic accretion events after the gap has formed. 
In \fig{fig:burstcomp}, we compare the accretion rate of the models `$\mathrm{low\_alpha}$' (black line) and `$\mathrm{low\_Bstar}$' (red line) with our reference model (blue line).
To compare the influence of the parameters on the episodic accretion events, we have shifted the time axis to the beginning of the first accretion event after the gap has formed. 
For both models, the characteristics of the accretion events are summarised in \tab{tab:burst_chara}.

For a low value of $\alpha_\mathrm{MRI}$ (model `$\mathrm{low\_alpha}$'), the piled-up material in the dead zone is accreted to the star at a slower rate compared to the reference model once the accretion event is triggered.
As a result, the maximum accretion rate is smaller and the accretion event lasts longer.
Once the accretion event is over, it takes a relatively long time until enough material has accumulated in the dead zone until the next accretion event is triggered and the frequency of the accretion events for the model `$\mathrm{low\_alpha}$' is smaller compared to the reference values.
Due to the lower viscosity parameter, the disk is more massive, the time span in which an accretion event is triggered is longer, and the number of accretion events is larger compared to the reference model.
The values of $R_\mathrm{gap}(\tau_\mathrm{eae})$ and $f_\mathrm{eae}$ have changed only slightly.

In case of a smaller value of the stellar magnetic field, $B_\star = 0.5$~kG (model `$\mathrm{low\_Bstar}$'), the inner disk boundary is located closer to the star (see \equo{eq:truncation_radius}) and the inner disk is hotter compared to the reference model.
Once an accretion event is triggered, a larger part of the inner disk becomes MRI active, and more mass is accreted on the star.
As a result, the accretion event lasts longer and it takes more time to replenish the inner dead zone with enough material to trigger the next accretion event.
Consequently, the number of accretion events is reduced compared to the reference model.
The value of $f_\mathrm{eae}$, on the other hand, is increased to 0.05.

\begin{table}[ht]
\centering
\caption{Characteristics of the episodic accretion events. The hash symbol ($\#$) denotes the number of episodic accretion events. The unit of the X-ray luminosities is erg per second (erg/s).}              
\begin{tabular}{l c c c c}         
\hline\hline 
Model & \# & $\tau_\mathrm{eae}$ [Myr] & $R_\mathrm{gap}(\tau_\mathrm{eae})$ [AU] & $f_\mathrm{eae}$ \\

\hline

no X-ray & 660 & 1.90 & --- & --- \\
$L_\mathrm{X}=10^{29}$ & 315 & 0.99 & --- & --- \\
$L_\mathrm{X}=10^{30}$ & 101 & 0.34 & 11.8 & 0.002 \\
$L_\mathrm{X}=10^{31}$ & 55 & 0.18 & 32.7 & 0.03 \\
$\mathrm{low\_alpha}$ & 68 & 0.32 & 28.6 & 0.03 \\
$\mathrm{low\_Bstar}$ & 46 & 0.17 & 29.0 & 0.05 \\

\hline\hline                                            
\end{tabular}
\label{tab:burst_chara}  
\end{table}

\subsection{Influence of photoevaporation on stellar rotation}\label{sec:PE_rot}

The initial stellar rotation period in all of our models discussed in the previous sections was set to six days.
Young star-forming regions, however, show stellar rotation periods mostly between one and ten~days \citep[e.g.][]{Rebull2006, Cody2010, Venuti2017, Rebull2020, Serna2021}{}{}.
We want to show how photoevaporation affects the evolution of the stellar rotation when starting from a fast or slow initial stellar rotation period with values of $P_\mathrm{init} = 1$~day and $P_\mathrm{init} = 10$~days, respectively.

\begin{figure}[ht!]
    \centering
         \resizebox{\hsize}{!}{\includegraphics{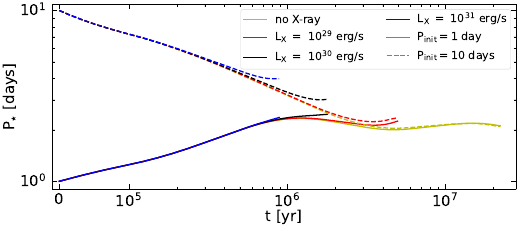}}
    \caption{
    Evolution of the stellar rotation period, $P_\star$, over time.
    The coloured lines correspond to the same X-ray luminosities as in \fig{fig:time}.
    The solid and dashed lines correspond to an initial stellar rotation period of $P_\mathrm{init} = 1$~day and $P_\mathrm{init} = 10$~days, respectively.
    }
    \label{fig:rotcomp}
\end{figure}

In \fig{fig:rotcomp}, we show the evolution of the stellar rotation period, $P_\star$, over time.
The solid and dashed lines correspond to an initial stellar rotation period of $P_\mathrm{init} = 1$~day and $P_\mathrm{init} = 10$~days, respectively.
In case no photoevaporation was included in the model (yellow lines), the rotation periods converge after $\approx 3$~Myr at a rotation period slightly above 2~days.
Afterwards, it is not possible to distinguish between the initially fast and slow rotating model.
This result agrees with \cite{Gehrig2023Paper4}.
The authors find that above an initial accretion rate of $\Dot{M}_\mathrm{rot} \sim 10^{-8}~\Msunpyr$, the initial distribution of rotation periods is forgotten and the stellar rotation period after the disk phase does not depend on its initial value but rather on the star-disk interaction.
With increasing X-ray luminosity the difference between the final stellar rotation periods in \fig{fig:rotcomp} increases.
The difference between a fast and slow initial rotation period is still small at the end of the simulation for $L_\mathrm{X} = 10^{29}~\mathrm{erg/s}$.
For $L_\mathrm{X} = 10^{31}~\mathrm{erg/s}$, the difference has increased to almost 2~days.

\begin{figure}[ht!]
    \centering
         \resizebox{\hsize}{!}{\includegraphics{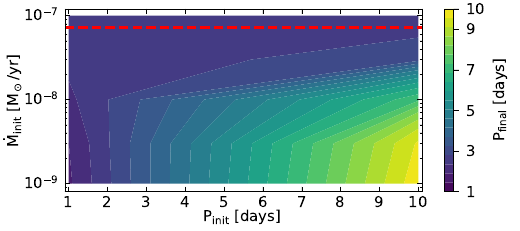}}
    \caption{
    Contour lines showing the final stellar rotation period, $P_\mathrm{final}$, after the simulation was completed for different initial accretion rates, $\Dot{M}_\mathrm{init}$, and initial stellar rotation periods, $P_\mathrm{init}$. 
    The X-ray luminosity for all models shown in this figure is $10^{30}~\mathrm{erg/s}$.
    The horizontal dashed line indicates the initial accretion rate, at which $P_\mathrm{final}$ differ by a maximum of 2\%.
    }
    \label{fig:contour}
\end{figure}

We wanted to explore how photoevaporation influences the magnitude of $\Dot{M}_\mathrm{rot}$ and varied the initial accretion rate, $\Dot{M}_\mathrm{init}$, of our model from $10^{-9}~\Msunpyr$ to $10^{-7}~\Msunpyr$ and the initial stellar rotation period, $P_\mathrm{init}$, from 1 to 10 days.
We note that initial accretion rates of $10^{-7}~\Msunpyr$ result in disk masses that are at the edge of being gravitationally unstable. Thus, we did not consider higher values for $\Dot{M}_\mathrm{init}$.
Similar to \cite{Gehrig2023Paper4}, the final rotation period, $P_\mathrm{final}$, after the simulation was completed for different initial accretion rates and rotation periods can be shown in a contour plot.
In \fig{fig:contour}, we show this contour plot exemplary for an X-ray luminosity of $L_\mathrm{X}= 10^{30}~\mathrm{erg/s}$.
The contour lines are almost vertical for low initial accretion rates of $\Dot{M}_\mathrm{init} \sim 10^{-9}~\Msunpyr$.
During the simulation time, the star-disk interaction does not significantly change the stellar rotation period and the initial distribution of rotation periods is still imprinted in the star after the disk phase.
When increasing the initial accretion rates, the contour lines become more and more horizontal.
The initial rotation periods become less important and the final rotation period after the disk phase depends on the star-disk interaction.
We have marked the initial accretion rate, at which $P_\mathrm{final}$ differ by a maximum of 2\%, with a horizontal dashed line, indicating $\Dot{M}_\mathrm{rot}$.

To quantify the `slope' of the contour lines in \fig{fig:contour}, we introduced a parameter, $C_\mathrm{rot}$, that compares the ratios of the initial stellar rotation periods and the rotation periods after the disk phase \citep[see][]{Gehrig2023Paper4}{}{},
\begin{equation}\label{eq:c_rot}
    \left( \frac{P_\mathrm{max}}{P_\mathrm{min}} \right)_\mathrm{final} = \left( \frac{P_\mathrm{max}}{P_\mathrm{min}} \right)^{C_\mathrm{rot}}_\mathrm{init} \, .
\end{equation}
A value of $C_\mathrm{rot}$ close to one corresponds to a weak star-disk interaction and the initial stellar rotation periods have not significantly changed during the star-disk interaction phase (nearly vertical contour lines in \fig{fig:contour} close to $\Dot{M}_\mathrm{init}\sim 10^{-9}~\Msunpyr$).
If the influence of the star-disk interaction increases and the initial rotational distribution is forgotten, the value of $C_\mathrm{rot}$ approaches zero, corresponding to nearly horizontal contour lines in \fig{fig:contour}.
A value of $C_\mathrm{rot}= 0.01$ corresponds to a maximum variation in the values of $P_\mathrm{final}$ of 2\% and indicates $\Dot{M}_\mathrm{rot}$.

\begin{figure}[ht!]
    \centering
         \resizebox{\hsize}{!}{\includegraphics{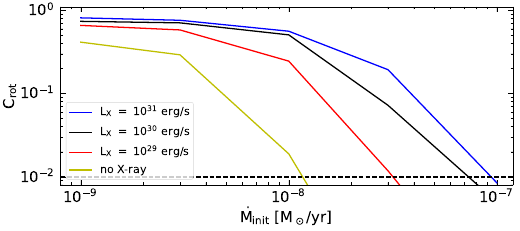}}
    \caption{
    Parameter $C_\mathrm{rot}$ for different initial accretion rates, $\Dot{M}_\mathrm{init}$.
    We use $C_\mathrm{rot}$ to compare the ratios of the initial stellar rotation periods and the rotation periods after the disk phase.
    The coloured lines correspond to the same X-ray luminosities as in \fig{fig:time}.
    The horizontal dashed line indicates a value of $C_\mathrm{rot}= 0.01$.
    }
    \label{fig:c_para}
\end{figure}

The relation between the initial accretion rate, $\Dot{M}_\mathrm{init}$, and $C_\mathrm{rot}$ is shown in \fig{fig:c_para}.
For all X-ray luminosities, the value is close to one for low initial accretion rates of $\Dot{M}_\mathrm{init}\sim 10^{-9}~\Msunpyr$.
In case no photoevaporation is included, the value of $C_\mathrm{rot}$ decreases fast towards $C_\mathrm{rot}=0.01$ slightly above $\Dot{M}_\mathrm{init}=10^{-8}~\Msunpyr$.
With increasing X-ray luminosities, the values of $\Dot{M}_\mathrm{init}$, at which $C_\mathrm{rot}=0.01$ increase almost up to $10^{-7}~\Msunpyr$ for an X-ray luminosity of $L_\mathrm{X}=10^{31}~\mathrm{erg/s}$.

\section{Discussion}
\label{sec:Discussion}

In this paper, we have extended a coupled star-disk evolution model \citep[][]{Steiner21, Gehrig2022, Gehrig2023Paper4} with the X-ray photoevaporation description of \cite{ercolano2021}.
With this model, we were able to calculate the effects of photoevaporation on the disk structure and evolution as well as on the stellar spin evolution self-consistently while including the innermost disk region ($< 0.1$~AU) and the effects of a stellar magnetic field.
Our results show that photoevaporation can affect the structure, evolution, and lifetime of the disk significantly.
Furthermore, our model enables the combined calculation of episodic accretion events and photoevaporation.
In the following, we compare our results to previous models and observations of transition disks and discuss possible implications regarding the formation and evolution of planets.

\subsection{Comparison with other models}

Over the past decade, several models have been developed to calculate the evolution of protoplanetary disks including the effects of photoevaporation.
In \cite{Bae2013PHOTO}, the evolution of a protoplanetary disk was studied including the early infall phase, episodic accretion events, and a relation between the X-ray luminosity and the stellar mass with X-ray luminosities of the order of $\sim 10^{30}~\mathrm{erg/s}$ after the infall phase.
Their values for the time in which the gap has been opened compared to the disk lifetime are larger by a factor of approximately 2 compared to our results, $f_\mathrm{gap}\lesssim 20$\%.
Similar to our findings, they conclude that opening a gap is difficult when considering low X-ray luminosities ($10^{29}~\mathrm{erg/s}$).
The biggest difference between the model presented in \cite{Bae2013PHOTO} and this study is the treatment of the inner disk boundary.
In their setup, \cite{Bae2013PHOTO} fixed the inner disk boundary to 0.1~AU and highlighted the importance of including the innermost disk region.
In our model, the variable inner disk boundary can approach the star even further (down to $\sim 0.01$~AU). 
In addition, we included the effects of stellar magnetic fields on the inner disk, the star-disk interaction, and a stellar spin model.
This allowed us to model the transition from an accreting disk to the propelling regime.
This process affects the evolution of the accretion rate significantly compared to a purely viscous evolution.
In the case of a viscous evolution, the accretion rate decreases slowly over time with a slope of $-1.5$ \citep[see Fig. 2 in ][]{Bae2013PHOTO} resulting in relatively long disk lifetimes of 2.5 to 4~Myr compared to our model with an X-ray luminosity of $10^{30}~\mathrm{erg/s}$.
We find that the accretion rate drops sharply when reaching the propelling regime (see \fig{fig:ref_mod}).
This difference could explain the larger values of $f_\mathrm{gap}$ in \cite{Bae2013PHOTO}.

Another model, we want to compare to our results is presented in \cite{Garate2021}.
They individually calculated the evolution of gas and dust including a dead zone and a photoevaporative mass-loss profile.
Similar to our setup, their initial disk corresponded to an early Class~II system and they studied the effects of X-ray luminosities ranging from $10^{30}~\mathrm{erg/s}$ to $10^{31}~\mathrm{erg/s}$.
The presence of a dead zone allows for the formation of large gaps while retaining accretion rates of $> 5\times 10^{-10}~\Msunpyr$.
The individual evolution of gas and dust enables the calculation of a radial dust profile and the generation of synthetic dust emission images that can be compared to observations.
There are, however, several important differences between the model presented in \cite{Garate2021} and this study.
First, the disk evolution model in \cite{Garate2021} was based on the simplified diffusion equation \citep[e.g.][]{LyndenBell74}{}{} and effects such as stellar magnetic torques or pressure gradients cannot be included.
Second, the inner boundary in \cite{Garate2021} was fixed to 1~AU.
As previously mentioned, the inner boundary is of significant importance for disk evolution as the star-disk interaction is strongest in the region $< 1$~AU.
Third, the transition from an accreting disk towards the propelling regime could not be included in \cite{Garate2021}, leading to longer disk lifetimes compared to this study.
Finally, the extent of the dead zone was fixed to a value of 5~AU, 10~AU, or 20~AU in \cite{Garate2021}, regardless of the disk's evolutionary state.
Our results, on the other hand, suggest \citep[similar to][]{Bae2013PHOTO}{}{} that the dead zone shrinks during the disk evolution with important implications for the formation of terrestrial planets (see \sref{sec:planets}).

In addition to the two highlighted models, there are other studies that calculate the disk evolution including photoevaporative profiles \citep[e.g.][]{Nakatani2018, Ercolano2018, Kunitomo2020}.
None of the previously mentioned models, however, combined the effects of the star-disk interaction with stellar magnetic torques and a stellar spin evolution, episodic accretion events, and a precise calculation of the radial dead zone boundaries.

\subsection{Observed high accretion rates in transition disks}

Over the past decade, a growing number of observed transition disks show a distribution of accretion rates that is comparable to classical T~Tauri stars with accretion rates up to $\gtrsim 10^{-7}~\Msunpyr$ \citep[e.g.][and references therein]{Cieza2012, Ercolano2017, Garate2021, Marel2023}.
While different photoevaporative disk evolution models can explain the observed distribution of gap sizes \citep[e.g.][]{Ercolano2018, Picogna2019, Garate2021}, high accretion rates of $\sim 10^{-7}~\Msunpyr$ cannot be explained \cite{Marel2023}. 
Based on the reviews of \cite{Ercolano2017} and \cite{Marel2023} the fraction of transition disks with high accretion rates of $\sim 10^{-7}~\Msunpyr$ is approximately 10\%.
However, none of the previous models consider the possibility of episodic accretion events. 
Instead, one of the prominent explanations for the observed high accretion rates in transition disks is a planetary companion \citep[e.g.][]{Lubow2006, Martel2022, Marel2023}.

\begin{figure}[ht!]
    \centering
         \resizebox{\hsize}{!}{\includegraphics{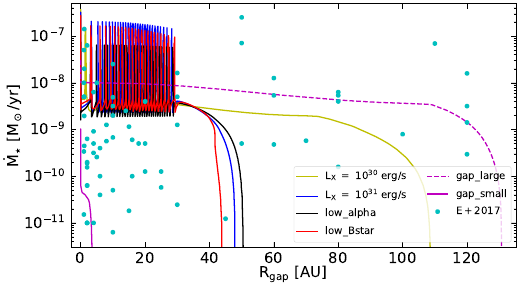}}
    \caption{
    Gap radius, $R_\mathrm{gap}$, with respect to the accretion rate, $\Dot{M}_\star$, (`$R_\mathrm{gap}\-- \Dot{M}_\star$'-wall) for the models from \tab{tab:burst_chara} that trigger accretion events.
    The cyan dots represent observational data taken from \cite{Ercolano2017} (labelled as `E+2017').
    To indicate the possible range of our models, we added two additional models: `$\mathrm{gap\_large}$' and `$\mathrm{gap\_small}$'.
    The models `$\mathrm{gap\_large}$' and `$\mathrm{gap\_small}$' have initial accretion rates of $\Dot{M}_\mathrm{init} = 3\times10^{-8}~\Msunpyr$ and $\Dot{M}_\mathrm{init} = 10^{-9}~\Msunpyr$ and X-ray luminosities of $10^{30}~\mathrm{erg/s}$ and $10^{29}~\mathrm{erg/s}$ were assumed, respectively.
    These models range from small accretion rates and small gap sizes in the lower-left corner to large gap sizes with accretion rates of $\lesssim 3\times 10^{-9}~\Msunpyr$ (lower-right side of the plot).
    }
    \label{fig:RgapMdot}
\end{figure}

For high X-ray luminosities, our model shows the possibility that such episodic accretion events can occur even after the gap has formed (see \fig{fig:timeBurst}).
The inner disk is sufficiently massive to increase the temperature in the dead zone above $T_\mathrm{active}$ and ignite accretion events with accretion rates of $\sim 10^{-7}~\Msunpyr$.
On the `$R_\mathrm{gap}\-- \Dot{M}_\star$'-wall these high accretion rates are located in the upper left corner (see \fig{fig:RgapMdot}). 
Previous studies have not been able to populate this region with high accretion rates and small gap radii with photoevaporative models \citep[e.g.][]{Garate2021}.
For reference, observational data taken from \cite{Ercolano2017} is shown as cyan circles.
Thus, our models are able to cover almost the entire `$R_\mathrm{gap}\-- \Dot{M}_\star$'-wall.
For an X-ray luminosity of $L_\mathrm{X} = 10^{31}~\mathrm{erg/s}$, the fractional time, $f_\mathrm{eae}$, in which the disk accretion rate is larger than $10^{-8}~\Msunpyr$ ranges between 3\% and 5\% of the time after the gap has formed (see \tab{tab:burst_chara}).
This value, however, drops fast for smaller X-ray luminosities.
This suggests that high accretion rates in transition disks could be explained without the presence of a planetary companion for strong X-ray luminosities.
Since $f_\mathrm{eae}$ is smaller compared to the fraction of strongly accreting transition disks and strongly dependent on $L_\mathrm{X}$, it is natural to assume that both mechanisms (companions and photoevaporation) are responsible for the observed accretion rates.

However, the distinction between these two mechanisms is difficult \citep[e.g.][]{Picogna2023}.
Observable dust features in protoplanetary disks or photoevaporative winds might reveal the nature of the gap \citep[e.g.][]{Franz2022, Picogna2023}.
Such features are not yet reproducible by our model as the evolution of dust is not yet included.
Comparing the relation between gap radius and accretion rates can also help distinguish between the gap opening mechanisms in case of high accretion rates.
The radius of the gap at the end of the episodic accretion events, $R_\mathrm{gap}(\tau_\mathrm{eae})$, has a value of approximately 30~AU. 
For larger gaps, the inner disk has been depleted to the extent that no additional accretion events can be triggered, and the accretion rates remain below $10^{-8}~\Msunpyr$.
Observations show a number of transition disks with high accretion rates of $> 10^{-8}~\Msunpyr$ and gap radii of larger than $\approx 40$~AU (see \fig{fig:RgapMdot}).
Such disk properties are not reproducible by our model and could be the result of a planetary or binary companion.
Clearly, the inclusion of dust evolution and the effects of a planetary companion in our model would provide additional insights into the characteristics of different gap-opening mechanisms.

\subsection{Implications for planet formation}\label{sec:planets}



We now want to discuss our results with respect to the formation of planets, especially small (terrestrial) planets that could provide the basis for life.
Over the past decades, it has become evident that terrestrial planets are likely to form in the inner regions ($< 10$~AU) of a protoplanetary disk where the dead zone is located \citep[e.g.][]{Matsumura2006, Mulders2018, Perotti2023}{}{}. 
The dead zone additionally provides protection against high-energy radiation that would ablate, for example, water which is a vital building block for life as we know it \citep[e.g.][]{Krijt2022, Perotti2023}.
Furthermore, planetary migration is significantly slowed down within the dead zone due to its low viscosity \citep[e.g.][]{Matsumura2006}.
Thus, the dead zone plays an important role in the formation, chemical evolution, and migration of planets.

Our model can track the extent of the dead zone during the disk evolution for different X-ray luminosities (see \fig{fig:time}).
For high X-ray luminosities, the extent of the dead zone drops towards 0.1~AU within 0.25~Myr, and the available time for a terrestrial planet to form is significantly reduced compared to low X-ray luminosities.
Even if a terrestrial planet has already formed in the shrinking dead zone, strong photoevaporation carries away the reservoir of water located in the dead zone \citep[][]{Perotti2023}, mitigating the chances of life on this planet.
To summarize, high X-ray luminosities and strong photoevaporation reduce the possibility of forming a terrestrial planet with access to a water reservoir in the dead zone.
It is interesting to note that our sun is considered to be among the weakest X-ray emitters among G~stars \citep[e.g.][]{Guedel2020}.

Another feature of the disk evolution dependent on the X-ray luminosity is the number and period of episodic accretion events (see \fig{fig:timeBurst}).
Each episodic accretion event changes the temperature structure of the protoplanetary disk \citep[e.g.][]{Steiner21}.
In addition, the stellar luminosity increases due to the enhanced accretion rate.
As a consequence, the chemical composition of the accretion disk changes during each accretion event \citep[e.g.][, and references therein]{Rab17}{}{}, affecting the composition of planets that are formed.

\subsection{Possible factors governing the stellar spin evolution}


The evolution of stellar rotation is strongly connected to other stellar parameters such as the emission of high energy radiation or the stellar magnetic field strength \citep[see the review of][]{Guedel2020}.
Fast-rotating stars show significantly higher levels of X-ray and EUV radiation and stronger magnetic field strengths compared to slow-rotating stars from stellar ages between $\sim 10$~Myr to several Gyr \citep[e.g.][]{Johnstone15b, Guedel2020, Johnstone2021}.
High-energy radiation can evaporate the atmosphere of a newly formed planet, which affects its habitability \citep[e.g.][]{Guedel97, Lammer03, Buccino06, Nortmann18}.
Stellar spin evolution models that include spin-dependent stellar wind formulations suggest that a star that rotates fast at the end of the accretion disk phase continues to be a fast rotator during its main sequence evolution \citep[e.g.][]{Johnstone15b, Johnstone2021}.
Consequently, fast-rotating stars provide unfavourable conditions for the evolution of a habitable planet.
Therefore, the question of what governs the stellar spin evolution during the disk phase should be raised.

In two recent studies, the parameters relevant to this question have been investigated. 
During the early phases of the star disk evolution (Class 0 and I systems), the stellar spin evolution depends strongly on the accretion history and the amount of energy added to the stellar interior by the accretion process \citep[][]{Gehrig2023Paper3}.
The stellar spin evolution during later evolutionary stages (Class II and III) depends more on the star-disk interaction \citep[][]{Gehrig2023Paper4}.
If the accretion rate at the beginning of the Class~II phase is of the order of $\gtrsim 10^{-8}~\Msunpyr$, the initial Class 0 and I evolution is completely forgotten and the stellar spin depends strongly on the stellar magnetic field strength and APSW.
If the accretion rate at the beginning of the Class~II phase is small, the importance of the star-disk interaction decreases, and the spin distribution of the Class~I systems remains unchanged during the disk evolution.

In this study, we have added the influence of photoevaporation to the stellar spin evolution of Class~II systems.
Confirming the results of \cite{Gehrig2023Paper4}, the initial spin distribution is forgotten without photoevaporation for initial accretion rates of $\Dot{M}_\mathrm{init} \sim 10^{-8}~\Msunpyr$.
When increasing the strength of the X-ray luminosity, this value increases up to $10^{-7}~\Msunpyr$ (see \fig{fig:c_para}).
Thus, only the spin evolution of massive Class~II disks with high accretion rates is governed by the star-disk interaction and their initial spin distribution is forgotten.
For smaller disks with lower accretion rates, the influence of the Class~I spin distribution on the final spin distribution after the disk phase becomes more relevant with increasing strength of photoevaporation.

\subsection{Stellar X-ray luminosity}
X-ray luminosities of classical T Tauri stars are observed to mainly be in the range of $10^{29}$ erg/s to $10^{31}$ erg/s \citep[e.g.][]{Gudel2007}{}{}, which motivates the restriction of X-ray photoevaporation models to these values. However, there are relations between the stellar X-ray luminosity and other stellar parameters such as bolometric luminosity, stellar mass, or accretion rate \citep[][]{Telleschi2006}{}{}. In our models with varying X-ray luminosity, we refrained from tracking these correlations and adapting the respective stellar or disk quantities. We did this in order to be able to isolate the effect a varying photoevaporation rate and corresponding mass-loss profile have on the evolution of the star-disk system. Changing other parameters, such as stellar mass or stellar bolometric luminosity, would lead to significant changes in stellar and disk structure, which would make the different models hardly comparable and would go beyond the scope of this study.

Observations show that during episodic accretion events, the X-ray output of the stellar system can increase significantly \citep[][]{Kastner2002, Kastner2004}{}{}. On the other hand, several studies indicate that there is a deficiency in stellar X-ray luminosity in strongly accreting T Tauri stars compared to their weakly accreting counterparts \citep[e.g.][]{Preibisch2005a, Telleschi2006, Bustamante2016}{}{}. Furthermore, the X-ray excess produced during an episodic accretion event could possibly be screened by dense accretion flows along magnetic funnels onto the star and therefore never reach the disk's surface. In order to accurately determine the exact profile of photoevaporative mass loss due to X-ray irradiation during an accretion event, the X-rays would have to be traced carefully. In our models, we did not aim to precisely track fluctuations in the photoevaporative mass loss. Additionally, the photoevaporation model constructed by \cite{Picogna2019} and expanded by \cite{ercolano2021} (which is implemented in our study) has been shown to saturate at high levels of stellar X-ray luminosities. Therefore, even if there was a strong, short-term X-ray excess during an episodic accretion event, the photoevaporative mass-loss rate would not be increased to a similarly large extent. The synthetic spectra calculated by \cite{ercolano2021}, which were used to determine the 1D photoevaporative mass-loss profiles, are based on observed X-ray spectra of about 500 T Tauri stars. The final spectral parameters were averaged over time based on
the large amount of observed X-ray sources. Therefore, the synthetic spectra already (at least partly) take into account short-term increases in luminosity due to the possibility of the observed objects having undergone episodic accretion events during the observation period. \\
The combination of these arguments led us to the decision to ignore the possible alteration of X-ray luminosity illuminating the disk during changes in the accretion rate on the star.

\subsection{Other sources of photoevaporation}
In the simulations conducted in this study, we included the influence of disk mass-loss due to heating and consequent escape of disk material by soft X-ray irradiation (0.1 ev $\lesssim~h\nu~\lesssim$ 1 keV) from the central star. However, radiation in the far ultraviolet
(FUV, $h\nu~\lesssim$ 13.6 eV) and extreme ultraviolet (EUV,  13.6 eV $\lesssim~h\nu~\lesssim~100$ eV) is capable of driving a photoevaporative wind as well. One of the differences between the heating by radiation of these different energy regimes is the ability of the photons to penetrate column densities, which is strongly wavelength dependent. Hereby, the penetration depth of EUV photons is on the order of one to two magnitudes smaller than the one of X-ray or FUV photons \citep[e.g.][]{Hollenbach2009}{}{}. This means EUV radiation only affects the particles in the uppermost layers of the inner parts of the disk and results in only small mass-loss rates of $\sim~10^{-10}~\Msunpyr$ \citep[][]{Hollenbach1994, Kunitomo2020}{}{}. Furthermore, when winds of other origin are driven from the disk (e.g. by X-ray or FUV heating), the EUV radiation is blocked by the wind columns and does not reach the disk's surface \citep[e.g.][]{Owen2012}{}{}. Therefore, heating by EUV radiation only plays a minor role in driving photoevaporative mass loss, especially in the presence of other wind-driving mechanisms.

FUV photons are not as easily absorbed and can penetrate further into the disk which can lead to rigorous mass-loss rates of $\sim10^{-8}~\Msunpyr$ \citep[e.g.][]{Gorti2009a, Wang2017, Nakatani2018}{}{}. Compared to the X-ray case included in our study, the mass-loss profile of FUV photoevaporation is shifted towards larger radii, enhancing the disk's mass-loss at distances $>100$ AU from the star. However, the heating by FUV radiation is very sensitive to the disk chemistry \citep[e.g.][]{Gorti2009a}{}{}, more specifically to the occurrence of PAHs and the dust distribution throughout the disk. This makes the effectiveness of FUV photoevaporation highly uncertain \citep[e.g.][]{Nakatani2018}{}{}. Furthermore, \cite{Owen2012} argue that in the presence of an X-ray-driven photoevaporative wind, FUV photoevaporation should only play a supporting role at radii smaller than $\sim 100$ AU. This relies on the fact that FUV photoevaporation penetrates deeper into the disk and highly depends on the disk structure and properties, while this is not the case for an X-ray-driven wind. 
Furthermore, heating by FUV radiation does not result in temperatures higher than a few $10^3$ K \citep[e.g.][]{Gorti2004}{}{}, while X-rays are capable of heating the disk material to temperatures of $>10^4$ K \citep[e.g.][]{Gorti2004, Ercolano2022, Pascucci2023}{}{}. So in the area of the disk where X-ray heating is effective, the sonic surface of the photoevaporative flow (which is where the mass-loss rate is evaluated) is located in the region predominantly heated by X-ray radiation \citep[][]{Owen2012}{}{}.
The mass flux at the sonic surface in an X-ray heated medium is purely determined by the stellar properties, meaning that the processes in the disk below the sonic surface have to result in the flow evaluated at the sonic surface. Therefore, at radii smaller than $\sim100$ AU, X-ray photoevaporation should dominate the mass loss. However, since the absorption of FUV photons is determined by the disk chemistry, this picture can change depending on the local penetration depth of FUV photons which is influenced by dust distribution and PAH abundance. 
Since disk chemistry and dust evolution were not included in our models, the inclusion of FUV photoevaporation was omitted. 

It is worth noting that a substantial fraction of the FUV luminosity originates from the accretion shock at the base of the funnel flows near the stellar surface and therefore strongly correlates with the accretion rate \citep[e.g.][]{Calvet2004}. 
The determination of the origin of EUV radiation is more difficult, but the accretion shock may deliver a significant contribution as well \citep[e.g.][]{Pascucci2014}.
Thus, during an event of enhanced accretion, the importance of FUV and EUV radiation for the photoevaporative mass-loss rate could increase. 
On the other hand, the accretion columns may be dense enough to screen the disk from EUV radiation originating from the accretion shock \citep[e.g.][]{Hollenbach2009}{}{}.
Furthermore, magnetospheric accretion features, such as funnel flows, seem to be absent in disks with high accretion rates \citep[e.g.][]{hartmann16}{}{}.
In those cases, the disk is pushed towards the stellar surface, and the energy released by the accretion process is blocked by the disk and advected into the star \citep[e.g.][]{hartmann11}{}{}. Although we cannot analyse this behaviour in our 1D approach, the contribution of FUV and EUV to the photoevaporation mass-loss would be affected as a consequence. 
A complete evaluation of the different processes, which depend on various stellar parameters as well as on the disk's chemical composition, is beyond the capabilities of our model. 

The models in our study represent the case of an isolated star-disk system, where photoevaporation is driven by the central star only. If the star is formed in an environment enriched with newly formed stars and corresponding radiation fields, external photoevaporation can influence the disk's evolution to a great extent \citep[e.g.][]{Scally2000, Adams2004a, Haworth2017}{}{}. This process is typically dominated by FUV photons irradiating the disk from the surrounding stellar population \citep[e.g.][]{Gorti2009a}{}{} and can cause photoevaporative mass-loss throughout the whole extent of the disk. Since the disk material is less gravitationally bound to the central star, external photoevaporation is most effective at the outer boundary of the disk, leading to a reduction of the disk mass as well as the disk radius. Depending on the density of the stellar population, the disk's lifetime can be reduced significantly by this process \citep[e.g.][]{Concha-Ramirez2021}{}{}. As a consequence, we expect that the timescales of the processes presented in our models would shorten and the relevance of the star-disk interaction for the stellar rotation period after the disk phase would decrease further if the effects of external photoevaporation were included. 

\subsection{Model limitations}

This study combines various crucial elements of stellar and disk evolution. However, the outcomes remain preliminary due to specific constraints within our model.

One-dimensionality: The main limitation of our current model is the restriction to the radial dimension.
Multi-dimensional models of the innermost disk region, including the dead zone and the star-disk interaction zone, are currently restricted to small timesteps and long-term calculations are not feasible \citep[e.g.][]{Steiner21}.
Several important, multi-dimensional processes are incorporated into our model in a simplified way.
This includes accretion via funnel flows, the calculation of the vertical and angular disk structure including non-axisymmetric features like planets, and the interaction with the stellar magnetic field.
A future version of our model is designed to extend to (at least) the vertical direction.

Magneto-centrifugal disk winds: In our model, the evolution of the accretion disk was calculated by combining viscous accretion and photoevaporation.
There is, however, another process that is assumed to have a significant impact on the accretion disk.
In the presence of a large-scale magnetic field, magneto-centrifugal winds can remove mass and angular momentum from the disk, driving accretion \citep[e.g.][]{Konigl11, Bai2013, Bai2015, Kunitomo2020}.
While photoevaporation has its strongest effects at radial distances of $\gtrsim 1$~AU (see \fig{fig:sigmadot}), magneto-centrifugal winds are also launched in the innermost disk region close to the star \citep[see the review of][]{Lesur2022}.
The angular momentum that is removed from the innermost disk region cannot be added to the star in the process of accretion and the stellar spin evolution is affected.
The interaction between stellar rotation, magneto-centrifugal winds, and photoevaporation could lead to new insights into the evolution and dispersal of protoplanetary disks.

Single stars: Currently, only disks around single, isolated stars are considered. 
Multiplicity, however, is common amongst young stars \citep[e.g.][]{Chen2013}.
A binary companion can also influence the evolution of the protoplanetary disk \citep[e.g.][]{Messina17, Zagaria2023}.
For example, the disk lifetime is significantly shorter in binary systems with separations below $\sim 100$~AU compared to single stars \citep[see review of][]{Offner2022}.
Furthermore, a close encounter with a stellar object can affect the star-disk system \citep[e.g.][]{Vorobyov17x}.
These effects are intrinsically non-axisymmetric and are therefore not included in the current version of our model.

Treatment of dust: The dust fraction in our model was kept constant over time and in the radial direction.
This simplification neglects important processes such as coagulation, settling, fragmentation, or drift. 
The dust fraction can vary significantly over time and for different disk radii \citep[e.g.][]{Testi2014, Vorobyov2022}.
In the context of transition disks, the dust properties at the gap boundaries can help distinguish between different gap-opening mechanisms \citep[planet or photoevaporation, e.g.,][]{Picogna2023}.
The inclusion of a dust evolution model will further improve our results and can be used to generate synthetic dust emission images.

Disk and stellar metallicity: 
In our model, we only used solar metallicity values for the star and the disk that are constant over time.
Different metallicity values, however, can affect the stellar parameters and the disk evolution \citep[e.g.][]{Amard19, Gehrig2023Paper2} and especially the photoevaporative mass-loss rates \citep[e.g.][]{Nakatani2018, Ercolano2018, Wolfer2019, Ercolano2010}.
The combination of our model with a sophisticated stellar evolution code that can include the effects of different metallicity values \citep[like in][]{Gehrig2023Paper2} could provide additional insights into the final stages of disk evolution.


\section{Conclusion}
\label{sec:conclusion}

In this study, we have combined a detailed disk model including the innermost disk regions, a stellar spin evolution model, and a model that describes photoevaporative mass loss in protoplanetary disks.
The stellar spin evolution, episodic accretion events, influence of stellar magnetic torques, properties of the dead zone, and formation and evolution of the photoevaporative gap can be calculated self-consistently in our models.
The main conclusions based on our results are as follows:

\begin{itemize}

    \item The dead zone is significantly affected by photoevaporation.
    High X-ray luminosities and the associated strong photoevaporative mass-loss rates deplete the dead zone within a very short time span ($\sim 0.2$~Myr) compared to low X-ray luminosities ($\sim 2$~Myr).
    As a consequence, the formation and evolution of (habitable) terrestrial planets in the dead zone are strongly impaired by high X-ray luminosities.

    \item The number and period of episodic accretion events decrease with increasing X-ray luminosity.
    Each accretion event can affect the chemical composition of the disk, and consequently, the composition of planets that are formed in the disk is subject to these processes as well.

    \item Even after the gap has formed, the inner disk contains enough material to trigger further episodic accretion events in case of high X-ray luminosities.
    High accretion rates of $\sim 10^{-7}~\Msunpyr$ in transition disks may not exclusively be the result of a planetary companion, but can also be explained by photoevaporation.

    \item Photoevaporation also affects the stellar spin evolution during the Class~II and Class~III evolution.
    In the case of no or only weak photoevaporation associated with low X-ray luminosities, the star-disk interaction strongly influences the stellar spin.
    For high X-ray luminosities, the influence of the star-disk interaction decreases, and the spin distribution of the previous Class~I phase prevails until the disk is dissolved.
    
\end{itemize}

\noindent  The results presented in this work highlight the effects of the interaction between long-term stellar and disk evolution with a particular focus on photoevaporative processes.
To understand young star-disk systems and their observed properties, a combined model, including different physical processes, can lead to new insight into their evolution.
An interesting prospect for future work is the interplay between photoevaporative- and magneto-centrifugal winds since magnetically driven outflows also affect the innermost disk regions and are capable of removing angular momentum. In combination with viscous angular momentum transport and stellar evolution, this will further deepen our understanding of the evolution of the star-disk system and the conditions governing the formation of planets.


\begin{acknowledgements}
We thank the anonymous referee for the constructive feedback that helped improve and clarify the manuscript.
\end{acknowledgements}


\bibliographystyle{resources/bibtex/aa}
\bibliography{literature/main}
%


\begin{appendix}
\section{Inclusion of the photoevaporation terms} \label{Appendix} 
The set of equations we solved (\equos{eq:cont}{eq:ene}) to conduct our simulations includes a term in every equation describing the mass-, momentum- and energy-loss, respectively, due to the photoevaporative process. 
The terms in the equations of motion and internal energy describe the respective loss of momentum and energy, which is held and consequently taken out of the disk system by the evaporating matter. 
Therefore, the terms in the equations of motion do not describe a torque or any kind of feedback that would, for example, slow down the bulge of the disk (which would be the case if considering magneto-hydrodynamic winds). 
Analogously, the term in the energy equation does not correspond to a cooling of the disk material. 
Omitting these terms would correspond to distributing the momentum and internal energy of the photoevaporating material in the disk before leaving the system.
To visualize the impact of these additional sink terms in \equos{eq:mot}{eq:ene}, we constructed a test case, in which we compared a model with all terms included (called `all~terms') to a model, in which the photoevaporative term was included only in the equation of continuity (called `only~cont'). The model parameters were taken from \tab{tab:model_para} and the stellar X-ray luminosity was set to $L_\mathrm{X} = 10^{31}~\mathrm{erg/s}$.
The initial model was a stationary disk without photoevaporation. 
To separate the influence of the photoevaporative terms from other effects, for example, pressure gradients close to the edges of the dead zone, we restricted the influence photoevaporation to a disk annulus between 10 and 20~AU, modelled as a Gaussian profile centred at 15~AU.

\begin{figure}[ht!]
    \centering
         \resizebox{\hsize}{!}{\includegraphics{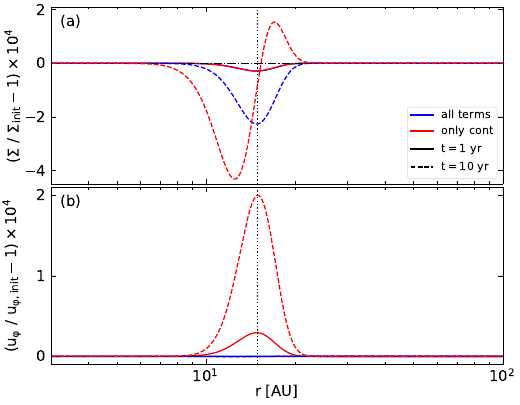}}
    \caption{Visualisation of the importance of photoevaporative terms in the equation of motion.
    The relative change of the surface density profile (Panel a) and the azimuthal velocity (Panel b) are shown for two different cases of the initial model.
    The coloured lines represent models in which the photoevaporative terms are included in all equations (blue lines, model `all terms') and only in the equation of continuity (red lines, model `only cont'), respectively.
    The solid and dashed lines show the relative changes one year and ten years after the photoevaporation has been enabled, respectively.
    No change compared to the initial model is indicated as a horizontal dash-dotted line at the value zero.}
    \label{fig:app}
\end{figure}

In \fig{fig:app}, we show the change of the surface density and azimuthal velocity of the disk after we enabled photoevaporation for simulation times of $t=1$~year (solid lines) and $t=10$~years (dashed lines) relative to the initial model.
The vertical dotted line indicates a radius of 15~AU, the centre of the annulus in which photoevaporation is active.
We note that the effect, which we would like to address, is reflected most clearly by visualizing the surface density and the azimuthal velocity of the disk, and thus, we restrict the following figure to these two quantities.

After one year (solid lines) both models show the expected sink in the surface density profile (see Panel a).
The profiles of the azimuthal velocity, on the other hand, already show a distinct difference after this short period.
While the angular velocity is unaffected by photoevaporation in the case of the Model `all~terms', we observe an acceleration in the case of Model `only~cont' (see Panel b).
Without the sink term in the equation of motion, the evaporated material deposits all its angular momentum in the remaining disk, and an excess of angular momentum starts to accumulate.
As time progresses (after ten years), Model `all~terms' behaves as expected.
The surface density sink deepens and the azimuthal velocity is nearly unaffected (blue dashed lines).
The growing surface density sink creates a pressure gradient that slightly affects the azimuthal velocity and causes a deviation from the initial model, which is barely noticeable.
The profile of the Model `only~cont' shows a significant difference after 10 years (red dashed lines).
The accumulating angular momentum excess and corresponding increase in azimuthal velocity affect the surface density profile.
Disk material is piled up near the outer edge of the region in which photoevaporation is active due to the outward force created by the excess of angular momentum.

This back reaction stands in contrast to what is expected from a purely photoevaporative model that removes disk material and does not affect the angular velocity of the remaining disk directly (compared, for example, to an MHD wind).
Thus, we included the photoevaporative sink terms in \equos{eq:cont}{eq:ene} to prevent this unwanted feedback.
We note, however, that this description assumes no interaction between the evaporated material and the surrounding disk.
An extension of our model in the vertical direction, for example, can certainly improve the current description.
\end{appendix}

\end{document}